\documentstyle[11pt,newpasp,twoside,epsf]{article}

\def\edcomment#1{\iffalse\marginpar{\raggedright\sl#1\/}\else\relax\fi}
\reversemarginpar

\begin{document}

\title{Stellar X-ray Binary Populations in Elliptical Galaxies}
\author{Raymond E. White III}
\affil{Department of Physics and Astronomy, University of Alabama, 
Tuscaloosa, AL 35487-0324}


\begin{abstract}
{\sl Chandra}'s high angular resolution can resolve emission from
stellar X-ray binaries out of the diffuse X-ray emission from gaseous
atmospheres within elliptical galaxies.  Variations in the X-ray binary
populations (per unit galaxian optical luminosity) are correlated with
variations in the specific frequency of globular clusters in ellipticals.
This indicates that X-ray binaries are largely formed in globular clusters,
rather than being a primordial field population.
\end{abstract}

\section{Introduction}

The X-ray emission from normal elliptical galaxies has two major
components: soft ($0.2-1$ keV) emission from diffuse gas and harder ($5-6$ keV)
emission from populations of accreting (low-mass) stellar X-ray binaries
(LMXB). The hardness of the LMXB component (placing its emission outside the
most responsive parts of the {\sl ROSAT} and {\sl ASCA} bandpasses) and its 
spatial confusion with the softer gaseous component have made it difficult to
constrain the global temperatures and luminosities of LMXB populations in
elliptical galaxies.  {\sl Chandra} observations are now resolving
out individual LMXBs in nearby ellipticals (Sarazin, Irwin \& Bregman 2000,
2001; Kraft et al.~2000; Finoguenov \& Jones 2001), making
their composite spectral analysis much easier.  Figure 1 compares
optical images of two ellipticals, NGC 1407 and NGC 4552, to their X-ray images
from the {\sl ROSAT} PSPC and {\sl Chandra} ACIS (White \& Davis 2001). 
The PSPC images emphasize the diffuse gaseous atmospheres of these ellipticals,
while the ACIS images are stretched to emphasize the discrete sources in each
galaxy.  {\sl Chandra} imaging has
clearly resolved out dozens of LMXBs from the diffuse gaseous emission in
these ellipticals, as it has in several other ellipticals in the work cited
above.

These recent {\sl Chandra} observations show that the hard X-ray emission
from normal ellipticals is dominated by LMXBs, not the advection-dominated
accretion flows (ADAFs) onto massive, central black holes, advocated by
Allen, di Mateo \& Fabian (2000).  
Meanwhile, as
{\sl Chandra} continues to observe more nearby ellipticals, there is a large
database of long {\sl ASCA} observations which has more to yield for ellipticals.

\begin{figure}[ht]
\caption{Comparison of optical and X-ray images ($5^\prime$ square) of the elliptical 
galaxies NGC 1407 and NGC 4552.  The {\sl ROSAT} PSPC images (middle) emphasize
X-ray emission from the diffuse gaseous atmospheres in each galaxy, while
the {\sl Chandra ACIS} images (right) are stretched to emphasize emission from
discrete sources (LMXBs).}
\vspace{3.5truein}
\plotfiddle{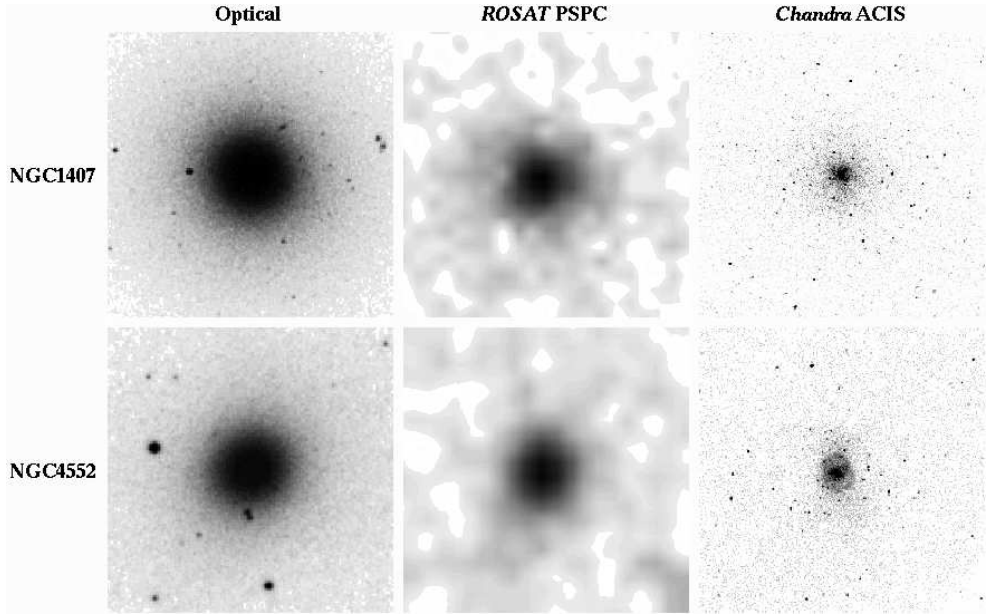}{0.0in}{0.0}{70}{70}{-210.0}{-130.0}
\end{figure}

\section{Constraining the LMXB Component in Ellipticals}

Strong spectral constraints on the hard stellar LMXB  component in
ellipticals can be made by simultaneously analyzing {\sl ASCA} spectra from
multiple ellipticals.   Most ellipticals require both soft (gaseous) and
hard (LMXB) components. I simultaneously fit {\sl ASCA} GIS spectra of six
ellipticals which provided individually reasonable spectral constraints on 
their hard emission. In the joint fits,  the temperature of the LMXB component 
was assumed to be the same for all galaxies; the temperatures of any soft
gaseous components (if present) were allowed to vary individually. The
resulting best-fit global spectrum of LMXBs  is fit equally well by a
bremsstrahlung spectrum with $kT = 6.3$ (5.2-7.9) keV or a
power-law spectrum with photon index $= 1.83$ (1.72-1.93),
where 90\% confidence limits are in parentheses. 
These are the tightest
constraints to date on the spectral properties of the stellar LMXB
component in ellipticals (White 2001) and are consistent with the spectral 
character of many individual LMXBs in our Galaxy.  
Fluxes for the LMXB components in an additional six ellipticals 
which had poorer photon statistics were determined by adopting the 
best fit LMXB temperature of 6.3 keV in fits to their GIS spectra.

In comparing the X-ray fluxes deduced for the LMXB component in these
ellipticals to the total optical magnitudes for these galaxies, I find 
there is a factor of ~4 range in the X-ray to optical flux ratio.
Although this range is much smaller than that of the softer gaseous
component (which has a factor of 100 range in X-ray/optical flux ratio), it
is still larger than expected, since the LMXB component is supposed to be
directly proportional to the stellar component. 
What is the source of the variance in the X-ray/optical flux ratio 
$f_{\rm LMXB}/f_{\rm opt}$?

\section{Globular Cluster Population Variations in Ellipticals}

Elliptical galaxies exhibit a wide range of globular cluster populations for
a given galaxian luminosity.  Furthermore, in our galaxy, LMXBs are produced
much more efficiently in globular clusters than in the field: globular
clusters contain $\sim$20\% of the known LMXBs, yet globular clusters contain
$<0.1$\% of the stars in our galaxy.  Apparently, globular clusters make LMXBs
$>200$ times more efficiently than the stellar field (Katz 1975), presumably
through tidal interactions between stars (Clark 1975).  It is therefore
conceivable that nearly ALL stellar LMXBs are formed in globular clusters
(Grindlay 1985; Kulkarni 2000). LMXBs which are not now in globular clusters
may have been ejected from globulars by supernova kicks immediately after
the primary collapsed to a neutron star. In this case, we might expect the
X-ray luminosity of the LMXB component to be correlated with the globular
cluster population of a galaxy, regardless of whether all LMXBs are
currently resident in globular clusters.  

\begin{figure}[ht]
\caption{The specific frequency of globular clusters plotted against
the ratio of global LMXB X-ray fluxes to the optical magnitudes (fluxes) of the 
host ellipticals.}
\vspace{3.5truein}
\plotfiddle{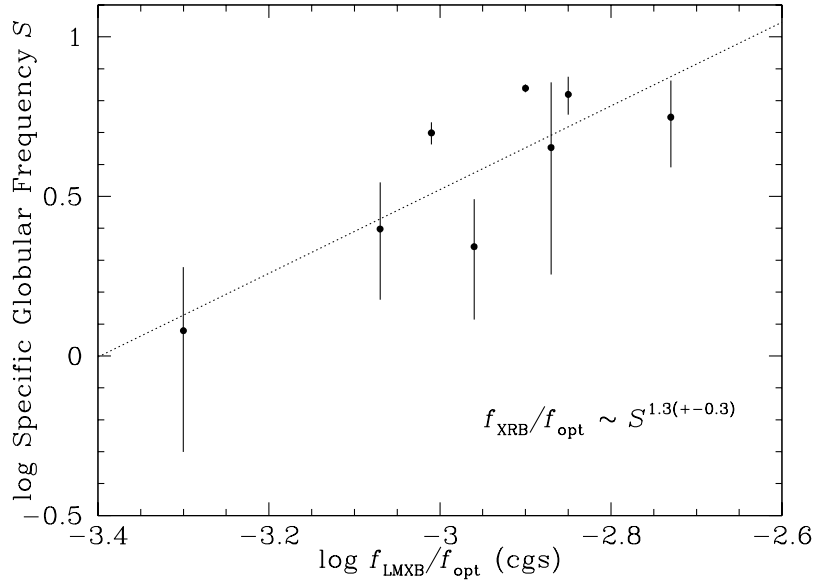}{0.0in}{-90.0}{40}{40}{-170.0}{280.0}
\end{figure}

To test this, I plot in
Figure 2 the specific frequency of globular clusters (the number of globular
clusters per unit galaxy luminosity, normalized by the luminosity corresponding
to a visual absolute magnitude of $M_V=-15$), 
versus the X-ray/optical flux ratio of the LMXB populations to the optical
flux (magnitude) of the host galaxies.
The globular cluster data are from the compilation of Kissler-Patig (1997),
while the X-ray data are from the {\sl ASCA} elliptical sample described
above (White 2001).  There appears to be a strong correlation, with the 
X-ray/optical flux ratio directly proportional (within the errors)
to the specific globular cluster frequency $S_{\rm gc}$:
$$f_{\rm LMXB} / f_{\rm opt} \propto S_{\rm gc}^{1.3\pm0.3}.$$
This strongly suggests that LMXB populations are indeed controlled by 
globular cluster populations.

\section{Conclusions}

Variations in the X-ray binary populations of elliptical galaxies
(per unit galaxian optical luminosity) are linearly
correlated with variations in their specific globular cluster frequencies.
This indicates that X-ray binaries are largely formed in globular clusters
(Grindlay 1985), rather than being a primordial field population. 
We predict that {\sl Chandra} observations of central dominant galaxies with
unusually large globular cluster populations will be found to have 
proportionally large numbers of LMXBs, as well.
For a more detailed analysis, see White, Kulkarni \& Sarazin (2001).

\end{document}